\documentclass[prd,aps,twocolumn,preprintnumbers, showpacs, nofootinbib,superscriptaddress,notitlepage]{revtex4-1}

\usepackage{amsmath,graphicx,epsfig,amssymb,dsfont,mathtools}
\usepackage[usenames]{color}
\usepackage{ulem} 
\usepackage{bigstrut}
\usepackage{slashed}
\usepackage{multirow}
\usepackage{subfigure}

\allowdisplaybreaks

\begin{document}

\title{Implications on CP violation of  charmless three body decays of bottom baryon from the U-spin analysis}

\author{Wei Wang} 
\affiliation{Shanghai Key Laboratory for Particle Physics and Cosmology,
 Key Laboratory for Particle Astrophysics and Cosmology (MOE),
 School of Physics and Astronomy, Shanghai Jiao Tong University, Shanghai 200240, China}
\affiliation{Southern Center for Nuclear-Science Theory (SCNT), Institute of Modern Physics, Chinese Academy of Sciences, Huizhou 516000, Guangdong Province, China}

\author{Zhi-Peng Xing}
\email{zpxing@nnu.edu.cn}
\affiliation{Department of Physics and Institute of Theoretical Physics, Nanjing Normal University, Nanjing, Jiangsu 210023, China}

\author{Zhen-Xing Zhao}
\email{zhaozx19@imu.edu.cn}
\affiliation{School of Physical Science and Technology, Inner Mongolia University, Hohhot 010021, China}

\begin{abstract}
Motivated by recent LHCb measurements of CP violation in $\Lambda_b$ three-body decays, we conduct an analysis of CP asymmetry in three-body decays of bottom baryons utilizing U-spin symmetry. We first develop a convenient representation  that  facilitates the incorporation of U-spin symmetry into our analysis. By integrating weak and strong phases into the derived amplitude, we obtain the CP asymmetry and establish relationships between CP asymmetries in U-spin related decay channels. With the help of experimental  measurements, we provide numerical predictions for the CP asymmetries for other decay channels, particularly $\Lambda_b \to \Sigma^0 K^+ K^-$, $\Xi^0_b \to \Lambda \pi^+ \pi^-$, $\Lambda_b \to \Sigma^0 K^+ \pi^-$, and $\Xi^0_b \to \Sigma K^- \pi^+$.  Furthermore, we provide a quantity that can be used as a null test of standard model and deviations from 0 will reflect the possible new physics.  These results are valuable for CPV measurements in baryon decays, and can be tested in future experiments.
\end{abstract}

\maketitle

\section{Introduction}


In heavy flavor physics, the study of CP violation (CPV) is undoubtedly one of the most important topics. It serves not only to account for the baryon-antibaryon asymmetry within the universe but also plays a vital role in unraveling the behaviors of QCD and electroweak  interactions in hadron decays. In the standard model of particle physics, CP violation in hadron decays is elucidated through the interplay of at least two  weak and strong phases within the decay amplitude. Consequently, delving into CP violation not only serves as a litmus test for the CKM matrix and the genesis of the weak phase but also propels our comprehension of QCD forward.

Although the CPV is of great importance, its study remains incomplete. Over the past six decades, CPV has been measured in K~\cite{Christenson:1964fg}, B~\cite{BaBar:2001ags,Belle:2001zzw}, and D~\cite{LHCb:2019hro} mesons. While these CPV effects are presumably  consistent with the standard model (SM) predictions, many CPV effects in hadron decay channels remain to be explored~\cite{CDF:2011ubb,CDF:2014pzb,Wang:2022hop,Wang:2022tcm,Geng:2022osc,Belle:2023str,Dai:2023zms,Song:2024jjn}, particularly in weak  decays of a heavy baryon. Very recently, the LHCb collaboration has reported the evidence for direct CPV measurement in the $\Lambda_b$ decay system~\cite{LHCb:2024yzj}: 
\begin{eqnarray}\label{LCPV}
&&A^{dir}_{CP}(\Lambda_b\to \Lambda \pi^+\pi^-)=-0.013\pm 0.051\pm0.018\notag\\
&&A^{dir}_{CP}(\Lambda_b\to \Lambda K^+ K^-)=0.083\pm0.023\pm0.016,\notag\\ 
&&A^{dir}_{CP}(\Lambda_b\to \Lambda K^+ \pi^-)=-0.118\pm0.054\pm0.021,\notag\\
&&A^{dir}_{CP}(\Xi_b^0\to\Lambda K^-\pi^+)=0.27\pm0.12\pm0.05.
\end{eqnarray} 
These exciting results motivate further theoretical studies.

In recent years, different theoretical studies have been conducted  on weak decays of  heavy  baryons  using approaches such as perturbative QCD~\cite{Lu:2009cm,Rui:2022jff,Zhang:2022iun,Rui:2022sdc,Rui:2023fiz,Yu:2024cjd}, final-state interactions (FSI~\cite{Bediaga:2022sxw,Wang:2024oyi,Jia:2024pyb,He:2024pxh},  flavor SU(3) symmetry analysis~\cite{Gronau:2013mza,He:2015fsa,He:2018joe,Roy:2019cky,Wang:2019dls,Roy:2020nyx,Wang:2022nbm,Sun:2024mmk}, and others~\cite{Hsiao:2014mua,Zhu:2016bra,Dery:2022zkt,Wang:2022tcm,Shen:2023eln,Shen:2023nuw}. It should be noted that a precision calculation of CP violation in a dynamical approach requires not only the  knowledge of low energy inputs including the transition form factors and baryonic lightcone distribution amplitudes which can be   pursued from the lattice QCD for the lowest moments~\cite{Bali:2015ykx,RQCD:2019hps} and the complete shape distribution~\cite{Deng:2023csv,Han:2023xbl,Han:2023hgy,Han:2024ucv,Chu:2024vkn}, but also requests a systematic  understanding of various power corrections.  In our work, we employ the flavor symmetry  approach, which is regarded as a powerful and  model-independent tool in baryon decay studies. Traditionally, the SU(3)$_F$ symmetry, which reflects the symmetry among u, d, and s quarks, has been widely applied to baryon decay processes~\cite{He:2015fwa,Lu:2016ogy,Geng:2017mxn,Wang:2017azm,Dery:2020lbc,Roy:2020nyx,Hsiao:2021nsc,Huang:2021aqu}, and the predictions have been under examinations    by experimental measurements.

In the context of CPV, U-spin symmetry, which reflects the symmetry between d and s quarks, is also a powerful tool~\cite{Gronau:2000zy,Gronau:2013mda,Bhattacharya:2015uua,Grossman:2018ptn,Dery:2021mll,Gavrilova:2022hbx,Schacht:2022kuj}. Although U-spin symmetry may involve significant symmetry-breaking effects when predicting the absolute squared amplitudes of baryon decays compared to SU(3)$_F$ symmetry, these effects are less pronounced when predicting CPV. Since CPV originates from the weak phase in the CKM matrix, which describes the mixing of three generations of quarks, and the u quark is an up-type quark, whereas the d and s quarks are down-type quarks, the symmetry between u and d/s quarks does not affect the weak phase or CPV. Therefore, U-spin symmetry is particularly useful for studying CPV and has been extensively applied in CPV research of bottom hadron decay~\cite{Gronau:2000zy,Gronau:2013mda,Bhattacharya:2015uua,Grossman:2018ptn,Dery:2021mll,Gavrilova:2022hbx,Schacht:2022kuj}. The main motif of this work is to investigate the implications of recent experimental data on three-body decays of bottom baryons under the U-spin symmetry. 

The rest of this paper is organized as follows.  In Sec. II, the framework of U-spin symmetry is established, for which we construct a convenient representation. At the end, the relationships between decay amplitudes are derived. In Sec. III, we derive the CPV relationships by incorporating weak and strong phases into the amplitudes. Using the newly available CPV data, we predict the CPV in other three-body decay channels of bottom baryons. A brief conclusion is provided in the last section.
Some details are collected in the appendix.

\section{U-spin analysis of bottom baryon charmless three body decay amplitudes}

In this section we will first introduce a convenient representation to study the U-spin symmetry in heavy baryon decays. 

Under the U-spin symmetry, which reflects the $d$ and $s$ quark flavor symmetry, the anti-triplet bottom baryons can be classified by the decomposition of the SU(2) group as $2\otimes 2 = 1 \oplus 3$. The light baryon octet can be divided into three parts: $\{p, \Sigma^+\}$, $\{n, \Sigma^0, \Lambda^0, \Xi^0\}$, and $\{\Sigma^-, \Xi^-\}$, which correspond to a doublet, triplet, and doublet, respectively.  As for the meson octet, it can also be classified into a doublet, an anti-doublet, and a triplet.   As a consequence these baryons and mesons can be represented as matrices:
\begin{eqnarray}\label{ma}
B_{b}^1&=&(\Xi_b^-),\;B_{b}^2=(\Lambda_b^0,\Xi_b^0),\;B^{2}_1=(p,\Sigma^+),\notag\\
M^2&=&(\pi^-,K^-),\;M^{\bar 2}=(\pi^+,K^+),\;B^{2}_2=(\Sigma^-,\Xi^-),\notag\\
M^3&=&\begin{pmatrix}
\frac{\sqrt{3}\eta_8}{2\sqrt{2}}-\frac{\pi^0}{2\sqrt{2}}&K^0 \\
\bar K^0&\frac{\pi^0}{2\sqrt{2}}-\frac{\sqrt{3}\eta_8}{2\sqrt{2}} \\
\end{pmatrix},\notag\\
B^3&=&
\begin{pmatrix}
-\frac{\Sigma^0}{2\sqrt{2}}-\frac{\sqrt{3}\Lambda^0}{2\sqrt{2}}& n \\
\Xi^0&\frac{\Sigma^0}{2\sqrt{2}}+\frac{\sqrt{3}\Lambda^0}{2\sqrt{2}}\\
\end{pmatrix},
\end{eqnarray}
where the $\Sigma^-, \Xi^-$ can also be expressed by matrix $B^{ijk}_{S/A}$ as $B^{ijk}_{A}=\epsilon^{ij1}(B^2_2)^k$ and $B^{ijk}_{S}=\epsilon^{ki1}(B^2_2)^j+\epsilon^{kj1}(B^2_2)^i$ which is similar with the $SU(3)$ representation in Ref.~\cite{Wang:2024ztg}. The baryon triplet can be expressed as $(B^3)^{ij}=\epsilon^{ik}(B^3)^j_k$. The details of constructing the matrices are given in Appendix~\ref{sec:appendix}.

To describe the bottom baryon charmless three-body decays, the low-energy effective weak Hamiltonian should also be represented by a U-spin matrix. The Hamiltonian that describes the $\Delta S = 1$ or $\Delta S = 0$ B decays is~\cite{Buchalla:1995vs}
\begin{align}
{\cal H}_{eff}&=\frac{G_F}{\sqrt{2}}\bigg[V^*_{ub}V_{uq}\bigg(\sum^2_1 C_i Q_1^{uq}+\sum^{10}_3C_i Q^q_i\bigg)\notag\\
&+V^*_{cb}V_{cq}\bigg(\sum^2_1C_i Q^{cq}_i+\sum^{10}_3C_i Q^q_i\bigg)\bigg], \; q=d,s,
\end{align}
where $C_i$ are the Wilson coefficients and $Q_i$ are the four quark operators. Since we will focus on the U-spin symmetry of the Hamiltonian, the operators can be written by omitting their Dirac structure as
\begin{eqnarray}
&&Q_{1,2}^{qs}\sim \bar b q \bar q s,\;Q_{1,2}^{qd}\sim \bar b q \bar q d,\;Q^s_{3-6}\sim \bar b s \sum \bar q^\prime q^\prime,\notag\\
&&Q^d_{3-6}\sim \bar b d \sum \bar q^\prime q^\prime,\; Q^s_{7-10}\sim \bar b s \sum e_{q\prime} \bar q^\prime q^\prime,\notag\\
&&Q^d_{7-10}\sim \bar b d\sum e_{q^\prime} \bar q^\prime q^\prime, \;\;q=u,d,s.
\end{eqnarray}
Here $Q_{1,2}$ represent the current-current operators and $Q_{3-10}$ represent the penguin operators.
These operators represent a U-spin doublet, where the d components (up in U-spin) are ${\cal H}_{eff} = V^*_{ub} V_{ud} U^d + V^*_{cb} V_{cd} C^d$, and the s components (down in U-spin) are ${\cal H}_{eff} = V^*_{ub} V_{us} U^s + V^*_{cb} V_{cs} C^s$. The Hamiltonian can be written in U-spin as ${\cal H}_{eff} = H_u + H_c$, and $H_{u,c}$ form the doublet, respectively, as
\begin{eqnarray}
H_u=(V^*_{ub}V_{ud},V^*_{ub}V_{us}),\;\;H_c=(V^*_{cb}V_{cd},V^*_{cb}V_{cs}).
\end{eqnarray}

According to the U-spin symmetry,   charmless three body decays of a bottom baryon can then be divided into eight types as
\begin{eqnarray}\label{uspin}
&&B_b^2\to B^3 M^2 M^{\bar 2},\;B_b^2\to B^3 M^3 M^{3},\;B_b^2\to B^2_1 M^2 M^{3},\;\notag\\
&&B_b^2\to B^2_2 M^3 M^{\bar 2},\;B_b^1\to B^3 M^3 M^2,\;B_b^1\to B^2_2 M^2 M^{\bar 2},\notag\\
&&B_b^1\to B^2_2 M^3 M^{3},\;B_b^1\to B^2_1 M^2 M^2.
\end{eqnarray}
The corresponding  amplitudes of bottom baryon doublet decays can be built as
\begin{eqnarray}
{\cal A}_{32\bar{2}}^{2} = \sum_{q=u,c}&\bigg( &a_{q}(B_{b}^{2})_{i}(H_{q})_{j}(B^{3})^{ij}(M^{2})^{k}(M^{\bar{2}})_{k}\nonumber \\
 & + & b_{q}(B_{b}^{2})_{i}(H_{q})_{j}(B^{3})^{ik}(M^{2})^{j}(M^{\bar{2}})_{k}\nonumber \\
 & + & c_{q}(B_{b}^{2})_{i}(H_{q})_{j}(B^{3})^{jk}(M^{2})^{i}(M^{\bar{2}})_{k}\bigg),
\end{eqnarray}
\begin{eqnarray}
{\cal A}_{333}^{2} =\sum_{q=u,c}&\bigg( & d_q (B_{b}^2)_i (H_q)_j (B^3)^{ij} (M^3)^l_k (M^3)^k_l\notag\\
&+& e_q (B_{b}^2)_i (H_q)_j (B^3)^{ik} (M^3)^j_l (M^3)^l_k \notag\\
&+& f_q (B_{b}^2)_i (H_q)_j (B^3)^{jk} (M^3)^l_k (M^3)^i_l \notag\\
&+& g_q (B_{b}^2)_i (H_q)_j (B^3)^{kl} (M^3)^i_k (M^3)^j_l \notag\\
&+& h_q (B_{b}^2)_i (H_q)_j (B^3)^{ij} (M^3)^k_k (M^3)^l_l \notag\\
&+& i_q (B_{b}^2)_i (H_q)_j (B^3)^{jk} (M^3)^i_k (M^3)^l_l \notag\\
&+& j_q (B_{b}^2)_i (H_q)_j (B^3)^{ik} (M^3)^j_k (M^3)^l_l \bigg),
\end{eqnarray}
\begin{eqnarray}
{\cal A}_{232}^{2} =\sum_{q=u,c}&\bigg( & k_q (B_{b}^2)_i (H_q)_j (B^2_1)^{i} (M^3)^k_k (M^2)^j\notag\\
&+& l_q (B_{b}^2)_i (H_q)_j (B^2_1)^{j} (M^3)^k_k (M^2)^i\notag\\
&+& m_q (B_{b}^2)_i (H_q)_j (B^2_1)^{j} (M^3)i_k (M^2)^k\notag\\
&+& n_q (B_{b}^2)_i (H_q)_j (B^2_1)^{k} (M^3)^i_k (M^2)^j\notag\\
&+& o_q (B_{b}^2)_i (H_q)_j (B^2_1)^{k} (M^3)^j_k (M^2)^i\notag\\
&+& p_q (B_{b}^2)_i (H_q)_j (B^2_1)^{i} (M^3)^j_k (M^2)^k \bigg),
\end{eqnarray}
\begin{eqnarray}
{\cal A}_{2 3 \bar2}^{2} =\sum_{q=u,c}&\bigg( & q_q (B_{b}^2)_i (H_q)_j (B_A)^{ijk} (M^3)^l_k (M^{\bar{2}})_l\notag\\
&+& r_q (B_{b}^2)_i (H_q)_j (B_A)^{ijl} (M^3)^k_k (M^{\bar{2}})_l\notag\\
&+& s_q (B_{b}^2)_i (H_q)_j (B_A)^{ikl} (M^3)^j_k (M^{\bar{2}})_l\notag\\
&+& t_q (B_{b}^2)_i (H_q)_j (B_A)^{jkl} (M^3)^i_k (M^{\bar{2}})_l\notag\\
&+& u_q (B_{b}^2)_i (H_q)_j (B_S)^{ijk} (M^3)^l_k (M^{\bar{2}})_l\notag\\
&+& v_q (B_{b}^2)_i (H_q)_j (B_S)^{ijl} (M^3)^k_k (M^{\bar{2}})_l\notag\\
&+& w_q (B_{b}^2)_i (H_q)_j (B_S)^{ikl} (M^3)^j_k (M^{\bar{2}})_l\notag\\
&+& x_q (B_{b}^2)_i (H_q)_j (B_S)^{jkl} (M^3)^i_k (M^{\bar{2}})_l\bigg),\notag\\
\end{eqnarray}
As for the bottom baryon singlet decay process, the amplitudes can be constructed as
\begin{eqnarray}
{\cal A}^1_{332}=\sum_{q=u,c}&\bigg(& A_q(H_q)_i (B^3)^{ij}(M^3)^k_j (M^{\bar 2})_k\notag\\
&+&B_q (H_q)_i (B^3)^{ik}(M^3)^j_j (M^{\bar 2})_k\notag\\
&+&C_q (H_q)_i (B^3)^{jk}(M^3)^i_j (M^{\bar 2})_k\bigg),
\end{eqnarray}
\begin{eqnarray}
{\cal A}^1_{22\bar 2}=\sum_{q=u,c}&\bigg(& D_q(H_q)_i (B^2_1)^{i}(M^2)^j (M^{\bar 2})_j\notag\\
&+& E_q(H_q)_i (B^2_1)^{j}(M^2)^i (M^{\bar 2})_j\bigg),
\end{eqnarray}
\begin{eqnarray}
{\cal A}^1_{233}=\sum_{q=u,c}&\bigg(& F_q (H_q)_i (B^2_2)^{i}(M^3)^j_j (M^3)^k_k\notag\\
&+&G_q (H_q)_i (B^2_2)^{i}(M^3)^k_j (M^3)^j_k\notag\\
&+&H_q (H_q)_i (B^2_2)^{j}(M^3)^k_j (M^3)^i_k\notag\\
&+&I_q (H_q)_i (B^2_2)^{k}(M^3)^j_j (M^3)^i_k\bigg),
\end{eqnarray}
\begin{eqnarray}
{\cal A}^1_{22\bar 2}=\sum_{q=u,c} K_q (H_q)_i (B^2_1)^{j}(M^2)^i (M^{\bar 2})_j. 
\end{eqnarray}
By expanding these amplitudes, the U-spin amplitudes of charmless bottom baryon three body decays are given in Table.~\ref{table1}. For convenience, the processes which involve $\eta$ and $\pi^0$ will not be included in our work. 

\begin{table*}[htbp!]
\caption{The U-spin amplitudes  of  charmless  bottomed baryon three body decays. Here $\lambda_s^u=V_{ub}^*V_{us}$, $\lambda_d^u=V_{ub}^*V_{ud}$, $\lambda_s^c=V_{cb}^*V_{cs}$ and $\lambda_d^c=V_{cb}^*V_{cd}$. The $\sum_q$ means sum the index $q$ with $q=u,c$. }\label{table1}
\begin{tabular}{cccc}\hline\hline
channel &amplitude&channel &amplitude \\\hline 
$\Lambda_b\to \Lambda K^+K^-$ & $ \frac{\sqrt{3}}{2\sqrt{2}}\sum_q(\lambda^q_s(a_q+b_q))$&$\Lambda_b\to \Sigma^0 K^+K^-$ & $ \frac{1}{2\sqrt{2}}\sum_q(\lambda^q_s(a_q+b_q))$ \\\hline
$\Xi^0_b\to \Lambda \pi^+\pi^-$& $ \frac{\sqrt{3}}{2\sqrt{2}}\sum_q(\lambda^q_d(a_q+b_q))$&$\Xi^0_b\to \Sigma^0 \pi^+\pi^-$ & $ \frac{1}{2\sqrt{2}}\sum_q(\lambda^q_d(a_q+b_q))$\\\hline
$\Lambda_b\to\Lambda\pi^+\pi^-$&$ \frac{\sqrt{3}}{2\sqrt{2}}\sum_q(\lambda^q_s(a_q+c_q))$&$\Lambda_b\to\Sigma^0\pi^+\pi^-$&$ \frac{1}{2\sqrt{2}}\sum_q(\lambda^q_s(a_q+c_q))$\\\hline
$\Xi^0_b\to\Lambda K^+K^-$&$ \frac{\sqrt{3}}{2\sqrt{2}}\sum_q(\lambda^q_d(a_q+c_q))$&$\Xi^0_b\to\Sigma^0K^+K^-$&$ \frac{1}{2\sqrt{2}}\sum_q(\lambda^q_d(a_q+c_q))$\\\hline
$\Lambda_b\to n K^+K^-$&$\sum_q\lambda^q_d a_q$&$\Xi^0_b\to \Xi^0 \pi^+\pi^-$&$\sum_q\lambda^q_s a_q$\\\hline
$\Lambda_b\to n \pi^+K^-$&$\sum_q\lambda^q_s b_q$&$\Xi^0_b\to\Xi^0 K^+\pi^-$&$\sum_q\lambda^q_d b_q$\\\hline
$\Xi^0_b\to n \pi^+K^-$&$\sum_q\lambda^q_d c_q$&$\Lambda_b\to\Xi^0 K^+\pi^-$&$\sum_q\lambda^q_s c_q$\\\hline
$\Lambda_b\to\Lambda K^+\pi^-$&$\frac{\sqrt{3}}{2\sqrt{2}}\sum_q(\lambda^q_d(b_q+c_q))$&$\Lambda_b\to\Sigma^0 K^+\pi^-$&$\frac{1}{2\sqrt{2}}\sum_q(\lambda^q_d(b_q+c_q))$\\\hline
$\Xi^0_b\to\Sigma^0 \pi^+ K^-$&$\frac{1}{2\sqrt{2}}\sum_q(\lambda^q_s(b_q+c_q))$&$\Xi^0_b\to\Lambda \pi^+ K^-$&$\frac{\sqrt{3}}{2\sqrt{2}}\sum_q(\lambda^q_s(b_q+c_q))$\\\hline
$\Lambda_b\to n \pi^+\pi^-$&$\sum_q\lambda^q_d(a_q+b_q+c_q))$&$\Xi^0_b\to\Xi^0 K^+ K^-$&$\sum_q\lambda^q_s(a_q+b_q+c_q)+$\\\hline
\hline
$\Lambda_{b}\to nK^{0}\bar{K}^{0}$ & $\sum_q\lambda_{d}^{q}(2d_{q}+e_{q}+f_{q}))$ & $\Xi_{b}^{0}\to\Xi^{0}K^{0}\bar{K}^{0}$ & $\sum_q\lambda_{s}^{q}(2d_{q}+e_{q}+f_{q})$\tabularnewline
\hline 
$\Lambda_{b}\to\Xi^{0}K^{0}K^{0}$ & $\sum_q\lambda_{d}^{q}g_{q}$ & $\Xi_{b}^{0}\to n\bar{K}^{0}\bar{K}^{0}$ & $\sum_q\lambda_{s}^{q}g_{q}$\tabularnewline
\hline 
$\Xi_{b}^{0}\to\Lambda K^{0}\bar{K}^{0}$ & $\frac{\sqrt{3}}{2\sqrt{2}}\sum_q(\lambda_{d}^{q}(2d_{q}+e_{q}+f_{q}+g_{q}))$ & $\Lambda_{b}\to\Lambda K^{0}\bar{K}^{0}$ & $\frac{\sqrt{3}}{2\sqrt{2}}\sum_q(\lambda_{s}^{q}(2d_{q}+e_{q}+f_{q}+g_{q}))$\tabularnewline
\hline 
$\Xi_{b}^{0}\to\Sigma^{0}K^{0}\bar{K}^{0}$ & $\frac{1}{2\sqrt{2}}\sum_q(\lambda_{d}^{q}(2d_{q}+e_{q}+f_{q}+g_{q}))$ & $\Lambda_{b}\to\Sigma^{0}K^{0}\bar{K}^{0}$ & $\frac{1}{2\sqrt{2}}\sum_q(\lambda_{s}^{q}(2d_{q}+e_{q}+f_{q}+g_{q}))$\tabularnewline
\hline
\hline 
$\Lambda_{b}\to p\bar{K}^{0}\pi^{-}$ & $\sum_q\lambda_{s}^{q}(o_{q}+p_{q})$ & $\Xi_{b}^{0}\to\Sigma^{+}K^{0}K^{-}$ & $\sum_q\lambda_{d}^{q}(o_{q}+p_{q})$\tabularnewline
\hline 
$\Lambda_{b}\to pK^{0}K^{-}$ & $\sum_q\lambda_{d}^{q}(m_{q}+p_{q})$ & $\Xi_{b}^{0}\to\Sigma^{+}\bar{K}^{0}\pi^{-}$ & $\sum_q\lambda_{s}^{q}(m_{q}+p_{q})$\tabularnewline
\hline 
$\Lambda_{b}\to\Sigma^{+}K^{0}\pi^{-}$ & $\sum_q\lambda_{d}^{q}(n_{q}+o_{q})$ & $\Xi_{b}^{0}\to p\bar{K}^{0}K^{-}$ & $\sum_q\lambda_{s}^{q}(n_{q}+o_{q})$\tabularnewline
\hline 
$\Lambda_{b}\to\Sigma^{+}K^{0}K^{-}$ & $\sum_q\lambda_{s}^{q}(m_{q}+n_{q})$ & $\Xi_{b}^{0}\to p\bar{K}^{0}\pi^{-}$ & $\sum_q\lambda_{d}^{q}(m_{q}+n_{q})$\tabularnewline
\hline 
\hline 
$\Lambda_{b}\to\Xi^{-}K^{0}K^{+}$ & $\sum_q\lambda_{d}^{q}(t_{q}+s_{q}-w_q-x_q)$ & $\Xi_{b}^{0}\to\Sigma^{-}\bar{K}^{0}\pi^{+}$ & $-\sum_q\lambda_{s}^{q}(t_{q}+s_{q}-w_q-x_q)$\tabularnewline
\hline 
$\Lambda_{b}\to\Xi^{-}K^{0}\pi^{+}$ & $\sum_q\lambda_{s}^{q}(q_{q}-v_{q}+2x_q)$ & $\Xi_{b}^{0}\to\Sigma^{-}\bar{K}^{0}K^{+}$ & $-\sum_q\lambda_{d}^{q}(q_{q}-v_{q}+2x_q)$\tabularnewline
\hline 
$\Lambda_{b}\to\Sigma^{-}K^{0}\pi^{+}$ & $\sum_q\lambda_{d}^{q}(s_{q}+t_{q}-2u_q+w_q+x_q)$ & $\Xi_{b}^{0}\to\Xi^{-}K^{0}\pi^{+}$ & $-\sum_q\lambda_{d}^{q}(q_{q}+u_{q}-2w_q)$\tabularnewline
\hline 
$\Lambda_{b}\to\Sigma^{-}\bar{K}^{0}K^{+}$ & $\sum_q\lambda_{s}^{q}(q_{q}+u_{q}-2w_q)$ & $\Xi_{b}^{0}\to\Xi^{-}\bar{K}^{0}K^{+}$ & $-\sum_q\lambda_{s}^{q}(s_{q}+t_{q}-2u_q+w_q+x_q)$\tabularnewline
\hline
\hline 
$\Xi_{b}^{-}\rightarrow\Xi^{0}K^{-}K^{0}$ & $\sum_q A_{q}\lambda_{s}^{q}$ & $\Xi_{b}^{-}\rightarrow n\bar{K}^{0}\pi^{-}$ & $-\sum_q(A_{q}\lambda_{d}^{q})$\tabularnewline
\hline 
$\Xi_{b}^{-}\rightarrow nK^{-}\bar{K}^{0}$ & $\sum_q C_{q}\lambda_{s}^{q}$ & $\Xi_{b}^{-}\rightarrow\Xi^{0}K^{0}\pi^{-}$ & $-\sum_q(C_{q}\lambda_{d}^{q})$\tabularnewline
\hline 
$\Xi_{b}^{-}\rightarrow\Sigma^{0}K^{-}K^{0}$ & $\frac{1}{2\sqrt{2}}\sum_q(A_{q}+C_{q})\lambda_{d}^{q}$ & $\Xi_{b}^{-}\rightarrow\Sigma^{0}\bar{K}^{0}\pi^{-}$ & $-\frac{1}{2\sqrt{2}}\sum_q(A_{q}+C_{q})\lambda_{s}^{q}$\tabularnewline
\hline 
$\Xi_{b}^{-}\rightarrow\Lambda K^{0}K^{-}$ & $\frac{\sqrt{3}}{2\sqrt{2}}\sum_q(A_{q}+C_{q})\lambda_{d}^{q}$ & $\Xi_{b}^{-}\rightarrow\Lambda\bar{K}^{0}\pi^{-}$ & $-\frac{\sqrt{3}}{2\sqrt{2}}\sum_q(A_{q}+C_{q})\lambda_{s}^{q}$\tabularnewline
\hline 
\hline 
$\Xi_{b}^{-}\rightarrow\Xi^{-}\pi^{-}\pi^{+}$ & $\sum_q D_{q}\lambda_{s}^{q}$ & $\Xi_{b}^{-}\rightarrow\Sigma^{-}K^{-}K^{+}$ & $\sum_q D_{q}\lambda_{d}^{q}$\tabularnewline
\hline 
$\Xi_{b}^{-}\rightarrow\Xi^{-}K^{+}\pi^{-}$ & $\sum_q E_{q}\lambda_{d}^{q}$ & $\Xi_{b}^{-}\rightarrow\Sigma^{-}K^{-}\pi^{+}$ & $\sum_q E_{q}\lambda_{s}^{q}$\tabularnewline
\hline 
$\Xi_{b}^{-}\rightarrow\Sigma^{-}\pi^{-}\pi^{+}$ & $\sum_q(D_{q}+E_{q})\lambda_{d}^{q}$ & $\Xi_{b}^{-}\rightarrow\Xi^{-}K^{-}K^{+}$ & $\sum_q(D_{q}+E_{q})\lambda_{s}^{q}$\tabularnewline
\hline 
\hline 
$\Xi_{b}^{-}\rightarrow\Xi^{-}K^{0}\bar{K}^{0}$ & $\sum_q(2G_{q}+H_{q})\lambda_{s}^{q}$ & $\Xi_{b}^{-}\rightarrow\Sigma^{-}K^{0}\bar{K}^{0}$ & $\sum_q(2G_{q}+H_{q})\lambda_{d}^{q}$\tabularnewline
\hline 
\hline 
$\Xi_{b}^{-}\rightarrow pK^{-}K^{-}$ & $\sum_qK_{q}\lambda_{s}^{q}$ & $\Xi_{b}^{-}\rightarrow pK^{-}\pi^{-}$ & $\sum_q K_{q}\lambda_{d}^{q}$\tabularnewline
\hline 
$\Xi_{b}^{-}\rightarrow\Sigma^{+}K^{-}\pi^{-}$ & $-\sum_qK_{q}\lambda_{s}^{q}$ & $\Xi_{b}^{-}\rightarrow\Sigma^{+}\pi^{-}\pi^{-}$ & $-\sum_q K_{q}\lambda_{d}^{q}$\tabularnewline
\hline\hline
\end{tabular}
\end{table*}

Since the amplitude can be divided into two parts corresponding to the different CKM matrix elements: ${\cal A} = V_{ub}^* V_{uq} {\cal A}_u + V_{cb}^* V_{cq} {\cal A}_c$, one can directly derive the simple U-spin relations for each amplitude $A_{u/c}$. Taking $\Lambda_b \to \Lambda K^+ K^-$ as an example, we have
\begin{eqnarray}
{\cal A}_q(\Lambda_b\to \Lambda K^+K^-)&=&\sqrt{3}{\cal A}_q(\Lambda_b\to \Sigma^0 K^+K^-)\notag\\
&=&\sqrt{3}{\cal A}_q(\Xi^0_b\to \Sigma^0 \pi^+\pi^-)\notag\\
&=&{\cal A}_q(\Xi^0_b\to \Lambda \pi^+\pi^-),
\end{eqnarray}
 where $q=u,c$. Although the simple U-spin relations can be derived, these relations cannot be transformed into the ratio of decay widths.  This is because the ratio of $A_u$ and $A_c$ corresponding to different processes is different. Therefore, we cannot directly predict the branching ratio of these channels. However, these relations are powerful for predicting CPV, which will be discussed in the next section.

\section{CP violation relations under U-spin symmetry}

The directly CP violation of charmless bottom baryon three body decays  can  be expressed as
\begin{eqnarray}
A_{CP}^{dir}=\frac{|{\cal A}(B_b\to B MM)|^2-|{\cal A}(\bar B_b\to \bar B \bar M\bar M)|^2}{|{\cal A}(B_b\to B MM)|^2+|{\cal A}(\bar B_b\to \bar B \bar M\bar M)|^2}.
\end{eqnarray}
Since the weak phase comes from the CKM matrix elements, the amplitude can be written as two parts corresponding to different CKM matrix elements, each of which is complex. Then, the amplitude of $\Delta S = \pm 1$ processes can be written as
\begin{eqnarray}
{\cal A}&=&V_{ub}^*V_{us}{\cal A}_u+V_{cb}^*V_{cs}{\cal A}_c,\notag\\
\bar{\cal  A}&=&V_{ub}V^*_{us}{\cal A}_u+V_{cb}V^*_{cs}{\cal A}_c.
\end{eqnarray}
The amplitude of  $\Delta S=0$ processes can be written as
\begin{eqnarray}
{\cal A}&=&V_{ub}^*V_{ud}{\cal A}_u+V_{cb}^*V_{cd}{\cal A}_c,\notag\\
\bar{\cal  A}&=&V_{ub}V^*_{ud}{\cal A}_u+V_{cb}V^*_{cd}{\cal A}_c.
\end{eqnarray}
The squared amplitude is
\begin{eqnarray}
&&|{\cal A}(B_b\to B MM)|^2=|V_{ub}^*V_{uq}|^2|{\cal A}_u|^2+|V_{cb}^*V_{cq}|^2|{\cal A}_c|^2\notag\\
&&+2{\cal R}_e(V_{ub}^*V_{uq}{\cal A}_u \times V_{cb}V_{cq}^*{\cal A}_c^*),\notag\\
&&|{\cal A}(\bar B_b\to \bar B \bar M \bar M)|^2=|V_{ub}V_{uq}^*|^2|{\cal A}_u|^2+|V_{cb}V_{cq}^*|^2|{\cal A}_c|^2\notag\\
&&+2{\cal R}_e(V_{ub}V_{uq}^*{\cal A}_u \times V^*_{cb}V_{cq}{\cal A}_c^*),\; q=u,c.
\end{eqnarray}
The directly CP violation is
\begin{eqnarray}
A_{CP}^{dir}&=&\frac{4{\cal I}_m(V_{ub}^*V_{uq}V_{cb}V_{cq}^*){\cal I}_m({\cal A}_u{\cal A}^*_c)}{|{\cal A}(B_b\to B MM)|^2+|{\cal A}(\bar B_b\to \bar B \bar M\bar M)|^2},\notag\\
&&q=d,s.
\end{eqnarray}

With the help of the expression for direct CP violation and the U-spin relation, we can easily derive the CP violation relations. We take the relation between $\Lambda_b \to \Lambda K^+ K^-$ and $\Xi_b^0 \to \Lambda \pi^+ \pi^-$ as an example. We have
\begin{eqnarray}
&&|{\cal A}(\Lambda_b\to \Lambda K^+K^-)|^2-|{\cal A}(\bar \Lambda_b\to \bar \Lambda K^+K^-)|^2\notag\\
&&\quad\quad\quad\quad\quad=4{\cal I}_m(V_{ub}^*V_{us}V_{cb}V_{cs}^*){\cal I}_m({\cal A}_u{\cal A}^*_c),
\end{eqnarray}
where $A_{u/c}=\frac{\sqrt{3}}{2\sqrt{2}}(a_{u/c}+b_{u/c})$.
For the U-spin related processes $\Xi_b^0\to \Lambda \pi^+\pi^-$, we can also have
\begin{eqnarray}
&&|{\cal A}(\Xi_b^0\to \Lambda \pi^+\pi^-)|^2-|{\cal A}(\bar \Xi_b^0\to \bar \Lambda \pi^+\pi^-)|^2\notag\\
&&\quad\quad\quad\quad\quad=4{\cal I}_m(V_{ub}^*V_{ud}V_{cb}V_{cd}^*){\cal I}_m({\cal A}_u{\cal A}^*_c).
\end{eqnarray}
Following the CKM matrix unitarity relation  ${\cal I}_m(V_{ub}^*V_{us}V_{cb}V_{cs}^*)=-{\cal I}_m(V_{ub}^*V_{ud}V_{cb}V^*_{cd})$, we can directly give
\begin{eqnarray}
&|{\cal A}(\Lambda_b\to \Lambda K^+K^-)|^2-|{\cal A}(\bar \Lambda_b\to \bar \Lambda K^+K^-)|^2\notag\\
&=-(|{\cal A}(\Xi_b^0\to \Lambda \pi^+\pi^-)|^2-|{\cal A}(\bar \Xi_b^0\to \bar \Lambda \pi^+\pi^-)|^2).
\end{eqnarray}
Then the $A_{CP}^{dir}$ ratio of these channels is
\begin{eqnarray}
&&R(A_{CP}^{dir})=\frac{A_{CP}^{dir}(\Lambda_b\to \Lambda K^+K^-)}{A_{CP}^{dir}(\Xi_b^0\to \Lambda \pi^+\pi^-)}\notag\\
&&=-\frac{|{\cal A}(\Xi_b^0\to \Lambda \pi^+\pi^-)|^2+|{\cal A}(\bar \Xi_b^0\to \bar \Lambda \pi^+\pi^-)|^2}{|{\cal A}(\Lambda_b\to \Lambda K^+K^-)|^2+|{\cal A}(\bar \Lambda_b\to \bar \Lambda K^+K^-)|^2}\notag\\
&&=-\frac{{\cal B}(\Xi_b^0\to \Lambda \pi^+\pi^-)\cdot \tau(\Lambda_b)}{{\cal B}(\Lambda_b\to \Lambda K^+K^-)\cdot\tau(\Xi_b^0)} ,\label{CPV}
\end{eqnarray}
where we have used $|{\cal A}(\Lambda_b\to \Lambda K^+K^-)|^2+|{\cal A}(\bar \Lambda_b\to \bar \Lambda K^+K^-)|^2\sim 2|{\cal A}(\Lambda_b\to \Lambda K^+K^-)|^2 $.
Therefore the CPV relation can be derived as
\begin{eqnarray}
A_{CP}^{dir}(\Lambda_b\to \Lambda K^+K^-) &=&-A_{CP}^{dir}(\Xi_b^0\to \Lambda \pi^+\pi^-)\notag\\
&&\times{\cal R}(\frac{\Xi_b^0\to \Lambda \pi^+\pi^-}{\Lambda_b\to \Lambda K^+K^-}),
\end{eqnarray}
where a ratio is defined for convenience: 
\begin{eqnarray}
{\cal R}(\frac{B_b\to B MM}{B^\prime_b\to B^\prime M^\prime M^\prime})=\frac{{\cal B}(B_b\to B MM)\cdot \tau(B_b^\prime)}{{\cal B}(B^\prime_b\to B^\prime M^\prime M^\prime)\cdot\tau(B_b)}.
\end{eqnarray}

Since both $\Lambda$ and $\Sigma^0$ correspond to one triplet in Eq.~\eqref{ma}, there is only a factor of $\sqrt{3}$ difference between the amplitudes involving $\Lambda$ and $\Sigma^0$. This leads to the direct CPV relations for channels involving $\Lambda$ and $\Sigma^0$, such that $A_{CP}^{dir}(B_b \to \Lambda M M) = A_{CP}^{dir}(B_b \to \Sigma^0 M M)$.

After applying the same strategy as in Eq.~\eqref{CPV}, we can derive the CPV relations for bottom doublet decays as follows:
\begin{eqnarray}\label{cpv}
&&\frac{A_{CP}^{dir}(\Lambda_b\to \Lambda/\Sigma^0 K^+K^-)}{A_{CP}^{dir}(\Xi_b^0\to \Lambda/\Sigma^0 \pi^+\pi^-)}=-{\cal R}(\frac{\Xi_b^0\to \Lambda/\Sigma^0 \pi^+\pi^-}{\Lambda_b\to \Lambda/\Sigma^0 K^+K^-}),\notag\\
&&\frac{A_{CP}^{dir}(\Lambda_b\to \Lambda/\Sigma^0 \pi^+\pi^-)}{A_{CP}^{dir}(\Xi_b^0\to \Lambda/\Sigma^0  K^+ K^-)}=-{\cal R}(\frac{\Xi_b^0\to \Lambda/\Sigma^0 K^+ K^-}{\Lambda_b\to \Lambda/\Sigma^0 \pi^+\pi^-}),\notag\\
&&\frac{A_{CP}^{dir}(\Xi_b\to \Xi^0\pi^+\pi^-)}{A_{CP}^{dir}(\Lambda_b\to n K^+K^-)}=-{\cal R}(\frac{\Lambda_b\to n K^+K^-}{\Xi_b\to \Xi^0\pi^+\pi^-}),\notag\\
&& \frac{A_{CP}^{dir}(\Lambda_b\to n \pi^+ K^-)}{A_{CP}^{dir}(\Xi^0_b\to\Xi^0 \pi^- K^+)}=-{\cal R}(\frac{\Xi^0_b\to\Xi^0 \pi^+ K^-}{\Lambda_b\to n \pi^+ K^-}),\notag\\
&& \frac{A_{CP}^{dir}(\Lambda_b\to \Xi^0 \pi^- K^+)}{A_{CP}^{dir}(\Xi^0_b\to n \pi^+ K^-)}
= -{\cal R}(\frac{\Xi^0_b\to n \pi^+ K^-}{\Lambda_b\to \Xi^0 \pi^- K^+}),\notag\\
&&\frac{A_{CP}^{dir}(\Xi^0_b\to \Lambda/\Sigma^0 \pi^+K^-)}{A_{CP}^{dir}(\Lambda_b\to \Lambda/\Sigma^0 K^+\pi^-)}=-{\cal R}(\frac{\Lambda_b\to \Lambda/\Sigma^0 K^+\pi^-}{\Xi^0_b\to \Lambda/\Sigma^0 \pi^+K^-}),\notag\\
&&\frac{A_{CP}^{dir}(\Xi^0_b\to\Xi^0 K^+K^-)}{A_{CP}^{dir}(\Lambda_b\to n\pi^+\pi^-)}=-{\cal R}(\frac{\Lambda_b\to n\pi^+\pi^-}{\Xi^0_b\to\Xi^0 K^+K^-}),\notag\\
&&\frac{A_{CP}^{dir}(\Xi^0_b\to\Xi^0 K^0\bar K^0)}{A_{CP}^{dir}(\Lambda_b\to n K^0\bar K^0)}=-{\cal R}(\frac{\Lambda_b\to n K^0\bar K^0}{\Xi^0_b\to\Xi^0 K^0\bar K^0}),\notag\\
&&\frac{A_{CP}^{dir}(\Lambda_b\to p  \pi^-\bar K^0)}{A_{CP}^{dir}(\Xi^0_b\to \Sigma^+ K^- K^0)}=-{\cal R}(\frac{\Xi^0_b\to \Sigma^+ K^- K^0}{\Lambda_b\to p  \pi^-\bar K^0}),\notag\\
&&\frac{A_{CP}^{dir}(\Xi^0_b\to \Sigma^+ \pi^-\bar K^0)}{A_{CP}^{dir}(\Lambda_b\to p K^- K^0)}=-{\cal R}(\frac{\Lambda_b\to p K^- K^0}{\Xi^0_b\to \Sigma^+ \pi^-\bar K^0}),\notag\\
&&\frac{A_{CP}^{dir}(\Xi^0_b\to p K^-\bar K^0)}{A_{CP}^{dir}(\Lambda_b\to \Sigma^+ \pi^- K^0)}=-{\cal R}(\frac{\Lambda_b\to \Sigma^+ \pi^- K^0}{\Xi^0_b\to p K^-\bar K^0}),\notag\\
&&\frac{A_{CP}^{dir}(\Lambda_b\to \Sigma^+ K^- K^0)}{A_{CP}^{dir}(\Xi^0_b\to p \pi^- \bar K^0)}=-{\cal R}(\frac{\Xi^0_b\to p \pi^- \bar K^0}{\Lambda_b\to \Sigma^+ K^- K^0}),\notag\\
&&\frac{A_{CP}^{dir}(\Xi^0_b\to \Sigma^- \pi^+ \bar K^0)}{A_{CP}^{dir}(\Lambda_b\to \Xi^- K^+ K^0)}=-{\cal R}(\frac{\Lambda_b\to \Xi^- K^+ K^0}{\Xi^0_b\to \Sigma^- \pi^+  K^0}),\notag\\
&&\frac{A_{CP}^{dir}(\Lambda_b\to \Xi^- \pi^+ K^0)}{A_{CP}^{dir}(\Xi^0_b\to \Sigma^- K^+ \bar K^0)}=-{\cal R}(\frac{\Xi^0_b\to \Sigma^- K^+ \bar K^0}{\Lambda_b\to \Xi^- \pi^+ K^0}),\notag\\
&&\frac{A_{CP}^{dir}(\Lambda_b\to \Sigma^- K^+ \bar K^0)}{A_{CP}^{dir}(\Xi^0_b\to \Xi^- \pi^+  K^0)}=-{\cal R}(\frac{\Xi^0_b\to \Xi^- \pi^+  K^0}{\Lambda_b\to \Sigma^- K^+ \bar K^0}),\notag\\
&&\frac{A_{CP}^{dir}(\Xi^0_b\to \Xi^- K^+ \bar K^0)}{A_{CP}^{dir}(\Lambda_b\to \Sigma^- \pi^+  K^0)}=-{\cal R}(\frac{\Lambda_b\to \Sigma^- \pi^+  K^0}{\Xi^0_b\to \Xi^- K^+ \bar K^0}).
\end{eqnarray}

With decay amplitudes in Table.~\ref{table1}, one can also obtain the CPV relations of bottom baryon singlet  as
\begin{eqnarray}\label{cpv2}
&&\frac{A_{CP}^{dir}(\Xi^-_b\to \Xi^0 K^0K^-)}{A_{CP}^{dir}(\Xi^-_b\to n \bar K^0\pi^-)}=-{\cal R}(\frac{\Xi^-_b\to n \bar K^0\pi^-}{\Xi^-_b\to \Xi^0 K^0K^-}),\notag\\
&&\frac{A_{CP}^{dir}(\Xi^-_b\to n \bar K^0K^-)}{A_{CP}^{dir}(\Xi^-_b\to \Xi^0 K^0\pi^-)}=-{\cal R}(\frac{\Xi^-_b\to \Xi^0 K^0\pi^-}{\Xi^-_b\to n \bar K^0K^-}),\notag\\
&&\frac{A_{CP}^{dir}(\Xi^-_b\to \Lambda/\Sigma^0 \bar K^0 \pi^-)}{A_{CP}^{dir}(\Xi^-_b\to \Lambda/\Sigma^0 K^0 K^-)}=-{\cal R}(\frac{\Xi^-_b\to \Lambda/\Sigma^0 K^0 K^-}{\Xi^-_b\to \Lambda/\Sigma^0 \bar K^0 \pi^-}),\notag\\
&&\frac{A_{CP}^{dir}(\Xi^-_b\to\Xi^- \pi^-\pi^+)}{A_{CP}^{dir}(\Xi^-_b\to \Sigma^- K^+K^-)}=-{\cal R}(\frac{\Xi^-_b\to \Sigma^- K^+K^-}{\Xi^-_b\to\Xi^- \pi^-\pi^+}),\notag\\
&&\frac{A_{CP}^{dir}(\Xi^-_b\to\Sigma^- K^-\pi^+)}{A_{CP}^{dir}(\Xi^-_b\to \Xi^- K^+\pi^-)}=-{\cal R}(\frac{\Xi^-_b\to \Xi^- K^+\pi^-}{\Xi^-_b\to\Sigma^- K^-\pi^+}),\notag\\
&&\frac{A_{CP}^{dir}(\Xi^-_b\to\Xi^- K^-K^+)}{A_{CP}^{dir}(\Xi^-_b\to \Sigma^- \pi^+\pi^-)}=-{\cal R}(\frac{\Xi^-_b\to \Sigma^- \pi^+\pi^-}{\Xi^-_b\to\Xi^- K^-K^+}),\notag\\
&&\frac{A_{CP}^{dir}(\Xi^-_b\to\Xi^- K^0\bar K^0)}{A_{CP}^{dir}(\Xi^-_b\to \Sigma^- K^0\bar K^0)}=-{\cal R}(\frac{\Xi^-_b\to \Sigma^- K^0\bar K^0}{\Xi^-_b\to\Xi^- K^0\bar K^0}),\notag\\
&&\frac{A_{CP}^{dir}(\Xi^-_b\to\Sigma^- \pi^0\bar K^0)}{A_{CP}^{dir}(\Xi^-_b\to \Xi^- K^0\pi^0)}=-{\cal R}(\frac{\Xi^-_b\to \Xi^- K^0\pi^0}{\Xi^-_b\to\Sigma^- \pi^0\bar K^0}),\notag\\
&&\frac{A_{CP}^{dir}(\Xi^-_b\to p K^-K^-)}{A_{CP}^{dir}(\Xi^-_b\to \Sigma^+ \pi^-\pi^-)}=-{\cal R}(\frac{\Xi^-_b\to \Sigma^+ \pi^-\pi^-}{\Xi^-_b\to p K^-K^-}),\notag\\
&&A_{CP}^{dir}(\Xi^-_b\to p K^-K^-)=A_{CP}^{dir}(\Xi^-_b\to \Sigma^+ \pi^-K^-)\notag\\
&&A_{CP}^{dir}(\Xi^-_b\to \Sigma^+ \pi^-\pi^-)=A_{CP}^{dir}(\Xi^-_b\to p K^-\pi^-). 
\end{eqnarray}

Using the above relations and the measured CPV in Eq.~\eqref{LCPV}, one can make the prediction of CPV for other related decay channels. Using the lifetime~\cite{ParticleDataGroup:2024cfk}, we find  the CPV of the related processes is predicted as: 
\begin{eqnarray}\label{pre}
&&A^{dir}_{CP}(\Lambda_b\to \Sigma^0 K^+K^-)=0.083\pm 0.028,\notag\\
&&A^{dir}_{CP}(\Xi^0_b\to \Lambda \pi^+\pi^-)=-0.090\pm 0.051,\notag\\
&&A^{dir}_{CP}(\Lambda_b\to \Sigma^0 K^+\pi^-)=-0.118\pm 0.058,\notag\\
&&A^{dir}_{CP}(\Xi^0_b\to \Sigma K^-\pi^+)=0.27\pm 0.13,
\end{eqnarray}
where we used the latest measured branching ratio of $\Xi^0_b\to \Lambda \pi^+\pi^-$ and $\Xi^0_b\to \Lambda K^-\pi^+$ in Ref.~\cite{LHCb:2024yzj} as
\begin{eqnarray}
{\cal B}(\Xi^0_b\to \Lambda \pi^+\pi^-)&=&(11.0\pm4.8)\times 10^{-6}\notag\\
{\cal B}(\Xi^0_b\to \Lambda K^-\pi^+)&=&(10.4\pm4.1)\times 10^{-6}.
\end{eqnarray}
Since the branching ratio of $\Xi^0_b \to \Lambda \pi^+ \pi^-$ is also measured in Ref.~\cite{LHCb:2024yzj}, its CPV is expected to be measured in the future.

One can also notice that the CPV of both $\Xi^0_b \to \Lambda K^- \pi^+$ and $\Lambda_b \to \Lambda^0 K^+ \pi^-$ has been measured. Since these two channels have CPV relations in Eq.~\eqref{cpv}, we can use the experimental data to test this symmetry and search for possible new physics effects that could violate these relations. We can define difference $\Delta$ as
\begin{eqnarray}
\Delta=\frac{A_{CP}^{dir}(\Xi^0_b\to \Lambda \pi^+K^-)}{A_{CP}^{dir}(\Lambda_b\to \Lambda K^+\pi^-)}+{\cal R}(\frac{\Lambda_b\to \Lambda K^+\pi^-}{\Xi^0_b\to \Lambda \pi^+K^-}).
\end{eqnarray}
If the U-spin symmetry is strict, the standard model should predict $\Delta = 0$. Using the experimental data, we find 
\begin{eqnarray}
\Delta = -1.8 \pm 1.6. 
\end{eqnarray}
This indicates that the experimental measurements align with the Standard Model by approximately $1.1\sigma$,  given the substantial uncertainty in the measurement of $A_{CP}(\Xi^0_b \to \Lambda \pi^+ K^-)$. With improved accuracy in experimental data in the future, we can leverage $\Delta$ to explore potential new physics.

\section{Conclusions}

In this study, inspired by the recent LHCb measurement  we have examined the three-body decays of bottom baryons and their CP violation. U-spin symmetry, which highlights the symmetry between the d and s quarks, plays a crucial role in the analysis of CPV. As it exclusively involves down-type quarks, the phase information is directly reflected. We have introduced a practical representation  for investigating U-spin symmetry. By constructing the U-spin matrices in Eq.~\eqref{ma}, the three-body decays of bottom baryons are categorized into 8 parts as shown in Eq.~\eqref{uspin}. Within each part, the amplitude under U-spin symmetry is derived, and the corresponding expressions for each channel are provided in Table~\ref{table1}. Utilizing the simple symmetry and the breakdown of the amplitude into ${\cal A}_u$  and ${\cal A}_c$  by its Hamiltonian, we are able to establish relationships for each amplitude in the three-body decays of bottom baryons. By comparing the amplitudes under U-spin symmetry, we deduce the CPV relations outlined in Eq.~\eqref{cpv}.

Using the recent results from the LHCb measurement,  we have  predicted the CPV for the U-spin related channels.  The CPV of four channels: $\Lambda_b \to \Sigma^0 K^+ K^-$, $\Xi^0_b \to \Lambda \pi^+ \pi^-$, $\Lambda_b \to \Sigma^0 K^+ \pi^-$, and $\Xi^0_b \to \Sigma K^- \pi^+$ are possibly sizable as shown in Eq.~\eqref{pre} and we recommend the experimentalists to conduct an analysis in future.  Furthermore, we have provided a quantity that can be used as a null test of standard model and deviations from 0 will reflect the possible new physics. 

\section*{Acknowledgements}

We thank Prof. Ji-Bo He and Prof. Wen-Bin Qian  for useful discussions. This work  is supported by NSFC under grants No. 12065020,  12125503, 12375088, 12335003, 12405113, and 12335003. This work was partially supported by SJTU Kunpeng $\&$ Ascend Center of Excellence.

\begin{appendix}
\section{The construction of U-spin matrix}
\label{sec:appendix}

Since the construction of expression in $B_b$ $M^3$ $M^2$ $M^{\bar2}$ $B^2_1$ and  $B^2_2$ are straightforward. In this section, we will focus on the derivation of matrix $B^3$ which reflect the $\Sigma^0,\Lambda,n,\Xi^0$. We take the $\Sigma^0$ and $\Lambda$ as an example. The flavor of $\Lambda$ and $\Sigma^0$ is u,d,s. Under the u-spin symmetry, the wave function of u,s can be expressed as singlet and triplet. The triplet wave function is $(dd,\frac{1}{\sqrt{2}}(ds+sd),ss)$.
For constructing the baryon state, we need add the u quark and the spin information. In the first step, we can direct add the u quark and spin without any symmetry as
\begin{eqnarray}
&&\frac{1}{\sqrt{2}}(ds+sd)\frac{1}{\sqrt{2}}(\uparrow \downarrow+\downarrow\uparrow)\times u \uparrow,\notag\\
&&\frac{1}{\sqrt{2}}(ds+sd)\uparrow \uparrow \times u  \downarrow.
\end{eqnarray}
Since in the baryon state the flavor must be symmetry, the symmetry wave function can be constructed by the permutation operation as
\begin{eqnarray}\label{uspin}
\phi_3&=&-\frac{1}{\sqrt{3}}\frac{1}{2}[(ds+sd)(\uparrow \downarrow+\downarrow\uparrow),u\uparrow]_{p}\notag\\
&&+\frac{\sqrt{2}}{\sqrt{3}}\frac{1}{\sqrt{2}}[(ds+sd)\uparrow \downarrow+\downarrow\uparrow,u\downarrow]_p,
\end{eqnarray}
where $[ds,u]_p$ is the permutation operation which is $[ds,u]_p=\frac{1}{\sqrt{3}}(uds+dus+dsu)$.  After constructing the U-spin wave function, we need to express the state $\Lambda$ and $\Sigma^0$ by U-spin wave function. The $\Lambda$ and $\Sigma^0$ flavor wave function under SU(3)$_F$ are
\begin{eqnarray}\label{su3}
\Lambda&=&\frac{1}{\sqrt{12}}\big\{[u\uparrow d\downarrow s\uparrow-u\downarrow d\uparrow s\uparrow]+{\rm permutation}\big\}\notag\\
&=&\frac{1}{2}\bigg(-[(ds+sd)\uparrow\uparrow,u\downarrow]_p+[ds\downarrow\uparrow,u\uparrow]_p\notag\\
&&+[sd\uparrow\downarrow,u\uparrow]_p\bigg),\notag\\
\Sigma^0&=&\frac{1}{6}\big\{2u\uparrow d\uparrow s\downarrow-u\uparrow d\downarrow s\uparrow-u\downarrow d\uparrow s\uparrow\notag\\
&&+{\rm permutation}\big\}\notag\\
&=&\frac{1}{2\sqrt{3}}\bigg([(ds+sd)\uparrow\uparrow,u\downarrow]_p+2[ds\uparrow\downarrow,u\uparrow]_p\notag\\
&&+2[sd\downarrow\uparrow,u\uparrow]_p-[ds\downarrow\uparrow,u\uparrow]_p-[sd\uparrow\downarrow,u\uparrow]_p\bigg),
\end{eqnarray}
Comparing the wave function under U-spin and SU(3)$_F$ in Eq.\eqref{uspin} and Eq.\eqref{su3}, we can derive
\begin{eqnarray}
\phi_3=-\frac{1}{2}\Sigma^0-\frac{\sqrt{3}}{2}\Lambda.
\end{eqnarray}
The SU(2) representation of baryon triplet can be expressed as
\begin{eqnarray}
B^3&=&
\begin{pmatrix}
\phi_3&\sqrt{2}n \\
\sqrt{2}\Xi^0&-\phi_3\\
\end{pmatrix}.
\end{eqnarray}
The expression in Eq.~\eqref{ma} can be directly derived.
\end{appendix}

\end{document}